\documentclass[a4paper, 10pt, conference]{ieeeconf}      

\IEEEoverridecommandlockouts                              
\usepackage{graphics} 
\usepackage{epsfig} 
\usepackage{mathptmx} 
\usepackage{times} 
\usepackage{amsmath} 
\usepackage{amssymb}  
\usepackage{graphicx}
\usepackage[ansinew]{inputenc}
\usepackage{tikz}
\usepackage[none]{hyphenat}

\title{\LARGE \bf
On detecting determinism and nonlinearity in  microelectrode
recording signals: Approach based on non-stationary surrogate data
methods}

\author{D. Guarín-Lopez,
        A. Orozco-Gutierrez,
        E. Delgado-Trejos and E. Guijarro-Estelles
\thanks{D. Guarín-Lopez  and A. Orozco-Gutierrez  are with the Department
of Electrical Engineering, Universidad Tecnológica de Pereira,
Colombia. e-mail: dlguarin@utp.edu.co, aaog@utp.edu.co}%
\thanks{E. Delgado-Trejos is with the Research center at the Instituto Tecnológico Metropolitano, Medellín, Colombia. e-mail:edilsondelgado@itm.edu.co}%
\thanks{E. Guijarro-Estelles is with the department of Electrical Engineering,  Universidad Politecnica, Valencia, Spain. e-mail:eguijarro@eln.upv.es}}

\begin{document}
\tikzstyle{inicio} = [rectangle,draw,minimum height=1em, text
width=5em, text centered] \tikzstyle{bloque} =
[rectangle,draw,minimum height=0.1em, text width=5em,,text
height=0.000001em, text centered] \tikzstyle{blo} =
[rectangle,minimum height=1em, text width=5em, text centered]
\tikzstyle{linea} =[draw,-latex, very thin]

\maketitle \thispagestyle{empty} \pagestyle{empty}

\begin{abstract}
Two new surrogate methods, the Small Shuffle Surrogate (SSS) and the
Truncated Fourier Transform Surrogate (TFTS), have been proposed to
study whether there are some kind of dynamics in irregular
fluctuations and if so whether these dynamics are linear or not,
even if this fluctuations are modulated by long term trends. This
situation is theoretically incompatible with the assumption
underlying previously proposed surrogate methods. We apply the SSS
and TFTS methods to  microelectrode recording  (MER) signals from
different brain areas, in order to acquire a deeper understanding of
them. Through our methodology we conclude that the irregular
fluctuations in MER signals possess some
determinism.
\end{abstract}

\section{INTRODUCTION}\label{seg1}
Stereotactic deep brain stimulation is a widespread treatment for
different kinds of neurological diseases, especially motor
disorders, such as Parkinson's disease \cite{Xue}. In this
procedure, electrodes are permanently implanted in the patient's
subthalamic nucleus  (STNs). They emit signals that reduce the
effect of chronic hyperactivity of STN. This treatment is specially
suited for long term patients who
suffer from side effects of the medical treatment.\\
One crucial and difficult task for neurosurgeons is locating the
target brain area to place a neuro-stimulator. Due to differences
between the image based target and the position eventually reached,
the neurosurgeon defines the final position of the microelectrode
based on the sound and waveform of the  microelectrode recording
(MER) signal \cite{Aboy}. Since MER signals are time-dependent
\cite{Israel:2004},
the detection of each area becomes a very complex task and its accuracy depends on the surgeon's ability \cite{Xue}.\\
Recently, there has been a widespread interest in finding an
automatic way to identify a brain area based on MER signals. Many
approaches have been proposed (e.g., time domain analysis
\cite{artc6}, wavelet transform \cite{Gemmar}, Hilber-Huang
transform \cite{Xue,Pinzon}, nonlinear dynamics analysis
\cite{1372361}), but none of these  techniques are yet considered as a solution to the problem of automatic classification \cite{1372361}.\\
What are the characteristics of the underlying system that generate
MER signals? (i.e., it is deterministic or stochastic, linear or
nonlinear). The answer to this crucial unresolved question could
lead to the selection of the best technique for classification. The
stationary surrogate data method \cite{bookthree} has been used to
answer this question in biomedical signals \cite{Trejos,Small3}. The
basic approach is to select a pseudo stationary sub-segment of the
data, but in MER signals this is still an issue \cite{Aboy}.
\\The methodology we present here is based on novel non-stationary
surrogate data methods, with which we attempt to show that MER
signals are realizations of a deterministic non-stationary process,
contrary to what has been suggested \cite{Xue,Pinzon}.
\section{Materials and Methods}\label{seg2}
In this section we present our data base, briefly introduce the
basic ideas of the standard surrogate data method and describe the
SSS and TFTS methods.
\subsection{Database}
The  MER  signals  used  in  this  study are  from  the Polytechnic
University  of  Valencia (UPV) and the Technological University of
Pereira (UTP). The UPV database acquisition parameters were:
Sampling frequency 24 kHz, resolution 16-bits, and 240.000 samples.
The UTP acquisition parameters were: Sampling frequency 24 kHz,
resolution 16-bits, and 48.000 samples. Each signal was labeled by a
neurophysiologists. There are 92 segmentes of Thalamic Nucleus, 105
of Subthalamic Nucleus, 100 of Zona Incerta and 109 of Substantia
Nigra pars Reticulata. The surgeries were performed on five patients
in Valencia (Spain) using the acquisition equipment LEADPOINT-TM of
Medtronic and on five patients in Pereira (Colombia) using the
acquisition equipment ISIS-MER of Inomed.
\subsection{Surrogate data methods}\label{seg2-a}
The surrogate data methods test an observed time series against a
hierarchy of null hypotheses. The procedure can be described as
follows. One starts with an observed time series which is to be
tested against the null hypothesis of the surrogate data test. The
standard surrogate data repertoire provides algorithms to test
against the hypotheses of (i) independent and identically
distributed noise; (ii) linearly filtered noise; or (iii) a
monotonic nonlinear transformation of linearly filtered noise.
Algorithms for each of these three hypotheses generate an ensemble
of artificial time series data: the surrogate data. These surrogate
data sets are guaranteed to have both the properties associated with
the underlying null hypothesis and are also similar to the original
observed data. Now, one simply evokes whatever statistic is of
interest and compares the value of this statistic computed from the
data to the distribution of values elicited from the surrogates. If
the statistic value of the data deviates from that of the
surrogates, then the null hypothesis may be rejected. Otherwise, it
may not.\\The three standard surrogate algorithms are known in the
literature as (i) Random shuffle surrogates (RS); (ii) Random phase
surrogates (RP); and, (iii) Amplitude adjusted Fourier transform
(AAFT) surrogates. These techniques are linear surrogate methods.
Unfortunately, the application of the surrogate data method is
limited to stationary time series \cite{bookthree}; in fact when
this method is applied to non-stationary time series the results are
unreliable \cite{bookthree}. One possible solution is to split the
non-stationary signal into segments which could be considered nearly
stationary, find the surrogate for each segment and then put the
surrogates of each segments together. However, this procedure is not
applicable to data with sudden changes like jumps or spikes
\cite{bookthree}.
\subsection{Small Shuffle Surrogates}\label{seg2-b} Recently, T.
Nakamura and M. Small \cite{tomomichi} proposed a new surrogate data
method named Small-Shuffle Surrogate (SSS), the null hypothesis
addressed by this algorithm is that irregular fluctuations are
independently distributed random variables (i.e there is no short
term dynamics or determinism). The SSS method is essentially an
extension of the RS surrogate algorithm to non-stationary data. The
SSS method can be stated as follow. Let the original data be
$\{x_{t}\}$, let $\{i_{t}\}$ be the index of $\{x_{t}\}$. 1) Obtain
$\{i^{\ '}_{t}\}=\{i_{t}\}+\mathcal{N}(0,A^{2})$, 2) let $\{r_{t}\}$
be the index of the sorted $\{i^{\ '}_{t}\}$ and 3) obtain the
surrogate data $\{s_{t}\}$ from $s_{t}=x( r_{t}  )$.\\ In this way
local structures or correlations in irregular fluctuations are
destroyed and global behaviors are preserved. \\ After applying the
method to an extensive number of real and simulated signals T.
Nakamura and M. Small \cite{tomomichi} found that selecting $A=1$ is
a fairly good choice.
\subsection{Truncated Fourier Transform Surrogates}\label{seg2-b}
If the null hypothesis addressed by the SSS algorithm can be
rejected, the next question is whether these dynamics are linear or
nonlinear. In order to answer  this question in non stationary data,
T. Nakamura, M. Small and Y. Hirata \cite{tomomichi2} proposed the
truncated Fourier transform surrogate (TFTS) method. The null
hypothesis addressed by this algorithm is that irregular
fluctuations are generated by a non stationary linear noisy system.
The TFTS algorithm works by preserving the low frequency phases in
the Fourier transform, but randomizing the high frequency
components. The method presented here is an extension of the RS
algorithm. 1) Compute the complex Fourier Transform
$\{X_{\omega}\}_{\omega}$ of the original data $\{x_{t}\}$, 2)
generate random phases $\{\phi_{\omega}\}$ such that
$\{\phi_{\omega}\} \sim U(-\pi, \pi)$ if $\omega > f_{c}$, and
$\{\phi_{\omega}\} = 0$ if $\omega \leq f_{c}$ and 3) obtain the
surrogate by computing the inverse Fourier transform of the complex
series $\{X_{\omega}e^{\imath \phi_{\omega}}\}_{\omega}$.\\While all
phases are not randomized in this method it is possible to
discriminate between linearity and non-linearity because the
superposition principle is valid only for linear data. i.e., when
data are nonlinear, even if the power spectrum is preserved
completely, the inverse Fourier transform data
using randomized phases will exhibit a different dynamical behavior.\\
The surrogate data generated by this method are influenced primarily
by the choice of the cutoff frequency $f_{c}$. If $f_{c}$ is too
high, the TFTS data are almost identical to the original data. In
this case, even if there is nonlinearity in irregular fluctuations,
one may fail to detect nonlinearity. Conversely, if $f_{c}$ is too
low, the TFTS data are almost the same as the linear surrogate data
and the long-term trends are not preserved. In this case, even if
there is no nonlinearity in irregular fluctuations, one may  wrongly
judge otherwise. The method for selecting the correct value of
$f_{c}$ is presented in \cite{tomomichi2}.

\section{Detecting determinism and nonlinearity}\label{sec3}
\begin{figure}[t]
\centering
\begin{tikzpicture}[node distance = 0.1cm, auto]

  \node [inicio] (te1) {\scriptsize{Data acquisition}};
  \node [bloque, right of=te1, node distance=2.7cm] (te2) {\scriptsize{Preprocessing}};
  \node [bloque, right of=te2, node distance=2.7cm] (te3) {\scriptsize{Selection of the discriminant statistics}};
  \node [bloque, below of=te3, node distance=1.6cm] (te4) {\scriptsize{Determination of the shortest segment}};
  \node [bloque, left of=te4, node distance=2.7cm] (te4a) {\scriptsize{Determination of $f_{c}$}};
  \node [bloque, left of=te4a, node distance=2.7cm] (te5) {\scriptsize{Application of the SSS and the TFTS methods}};
  \node [bloque, below of=te5, node distance=1.8cm] (te6) {\scriptsize{Calculation of the statistics for the data and the surrogates}};
  \node [bloque, right of=te6, node distance=2.7cm] (te7) {\scriptsize{Acceptance or rejection of the null hypothesis}};

  \path [linea] (te1) -- (te2);
  \path [linea] (te2) -- (te3);
  \path [linea] (te3) -- (te4);
  \path [linea] (te4) -- (te4a);
  \path [linea] (te4a) -- (te5);
  \path [linea] (te5) -- (te6);
  \path [linea] (te6) -- (te7);
\end{tikzpicture}
\caption{General procedure of the proposed method in order to detect
determinism in MER signals. }\label{fig1}
\end{figure}
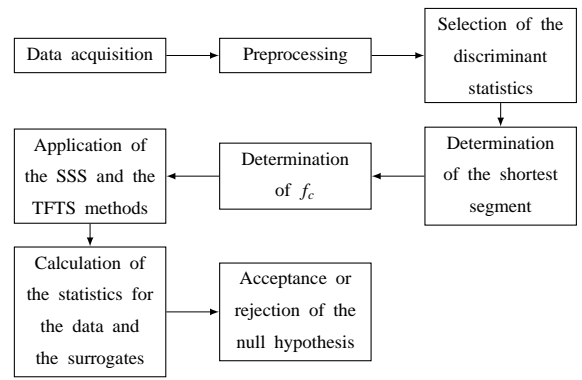

For the detection of determinism and nonlinearity in MER signals we
apply the SSS and TFTS methods respectively. The procedure
summarized in Fig. \ref{fig1}.
\subsection{Preprocessing}\label{sec3-a}
Prior to MER signal analysis, raw data is magnified by a
preamplifier located near the electrode to reduce electrical noise.
After these preconditioning steps, the signal is sampled with an
analog-to-digital converter with a sampling rate of  24 kHz. Then an
artefact detector is used to eliminate wrong entries in the MER
signal due to patient movement. Each MER signal is segmented with a
window of  1s, which is considered enough time for identification of
brain zones  \cite{Pinzon}.
\subsection{Selection of the discriminant statistics}\label{seg2-c}
 Dynamical measures are often used as discriminating statistics.
According to \cite{bookthree}, the correlation dimension  is one of
the most popular choices. To estimate these, we first need to
reconstruct the underlying attractor. For this purpose, a time-delay
embedding reconstruction is  usually applied \cite{bookthree}. But
this method is not useful for data exhibiting irregular fluctuations
and long-term trends. This is because a smaller time delay is
necessary to treat irregular fluctuations and a larger time delay is
necessary to treat long-term trends. At the moment, there is not a
good method for embedding
such data \cite{bookthree}.\\
Therefore, as discriminant statistics we chose the Average Mutual
Information (AMI) and the Lempel-Ziv Complexity (LZ Complexity) (see
\cite{bookthree} for further information). These are selected for
four reason: i) Using these statistics we avoid the difficulties
associated with embedding; ii) both are widely used in the
literature as discriminating statistics \cite{tomomichi2,Aboy2006},
iii) it has been shown that LZ complexity is suited to physiological
signals \cite{Aboy2006} and iv) in a separate study we conclude that
the obtained result using the correlation dimension as test
statistic are the same as when using AMI and LZ Complexity.
\subsection{Determination of the shortest segment to
analyze}\label{sec3-b} In order to determine the shortest segment to
analyze, we generated 24 sub-segments from the 1s segment, the first
sub-segment of 1000 data points (0.416s) and the last one of 24000
data points (1s), increasing 1000 data points each sub-segment. Then
we computed the AMI and LZ complexity for the 24 sub-segments.
\subsection{Determination of the correct value of
$f_{c}$}\label{sec3-c}
 In order to estimate the correct value
of $f_{c}$ we start with a high $f_{c}$ (i.e., we randomize the
phases of the highest $5\%$ of the frequency range; in this case the
frequency range is $1 - 12.000$ Hz due to the symmetry of the
Fourier coefficients), then if the auto correlation (AC) of the
original data falls within the distribution of the surrogates
generated with the TFTS algorithm (when the AC of the original data
falls within the distribution, linearity and long term trends are
sufficiently preserved in the surrogate data, we inspect the AC at
time lag 1 because it must be more sensitive to the nature of the
data \cite{tomomichi2}), we decreases the value of $f_{c}$ by a
constant rate (i.e., now we randomize the phases of the highest
$10\%$ of the frequency range). We keep doing this until we find a
value of $f_{c}$ for which the AC of the original data falls outside
the distribution of the surrogates, and then the correct value of
$f_{c}$ is the last one for which the AC of the original data fell
within the distribution of the surrogates.
\subsection{Application of the SSS and TFTS methods}\label{sec3-d}
In order to detect determinism and nonlinearity we generate 39
surrogates for each signal with each method, in this way for a two
sided test we achieve $95\%$ confidence, i.e., there is a $5\%$
probability that the null hypothesis is rejected even though it is
true.\\ Then we calculate the AMI at time lag 1 and the LZ
complexity for each signal and its surrogates, and check whether the
statistics for the original data falls within the distribution of
the surrogates. This information lead us to reject or not the null
hypothesis.
\section{Results and discussion}\label{sec4}
Following the procedure proposed in \ref{sec3-b}, we found that the
LZ-complexity and the AMI are well behaved for MER signals with
$n>2\times10^{4}$ (number of data points), so we decided to perform
all further analysis with the 1s window. Following \ref{sec3-c} we
found that by randomizing the phases of the highest $80\%$ of the
frequency range, the AC of the data fell outside the distribution of
the surrogates, so we decided to randomize the phases only of the
higher $75\%$ of the frequency range, in this case with a data
length of $24000$ data points we obtained $f_{c}=3000$ Hz.
\begin{figure}[t]
  \centering
  \includegraphics[width=7cm]{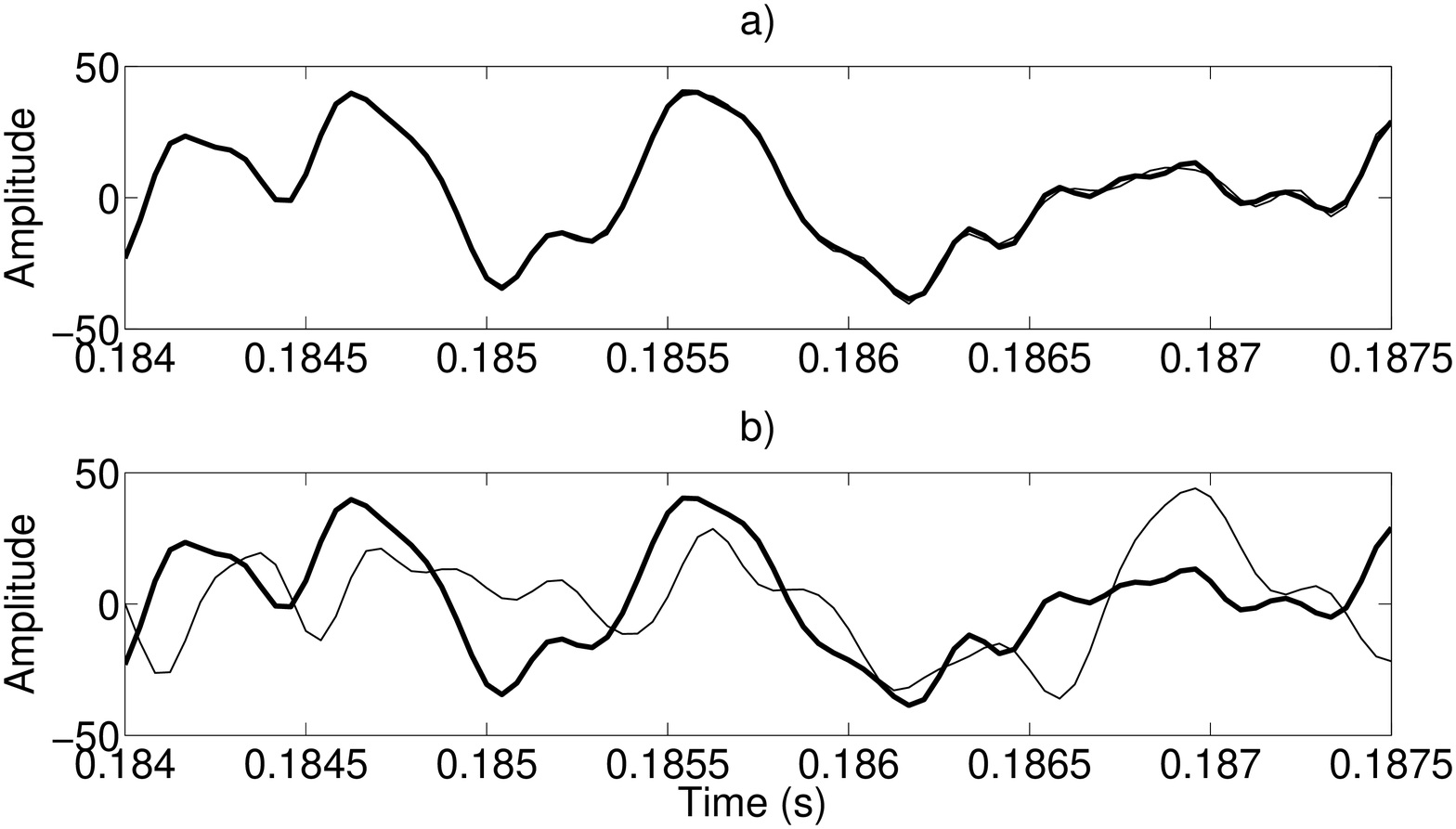}\\
  \caption{a) Original MER signal and one of the TFTS surrogates
  (thinner line) with $f_{c}=9600$ Hz b) Original MER signal and one
  of the TFTS surrogates (thinner line) with $f_{c}=1600$  Hz.}\label{fig2}
\end{figure}

Fig. \ref{fig2} a) shows the case of too little randomization, while
Fig. \ref{fig2} b) shows the case of too much randomization. In both
cases one could wrongly accept or reject a null hypothesis.
\begin{figure}[b]
  \centering
  \includegraphics[width=7cm]{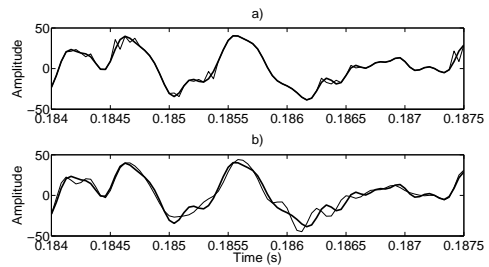}\\
  \caption{a) Original MER signal and one of the SSS surrogates
  (thinner line) b) Original MER signal and one of the TFTS
  surrogates with $f_{c}=7600$ (thinner line)}\label{fig3}
\end{figure}

\begin{figure}[t]
  \centering
  \includegraphics[width=7cm]{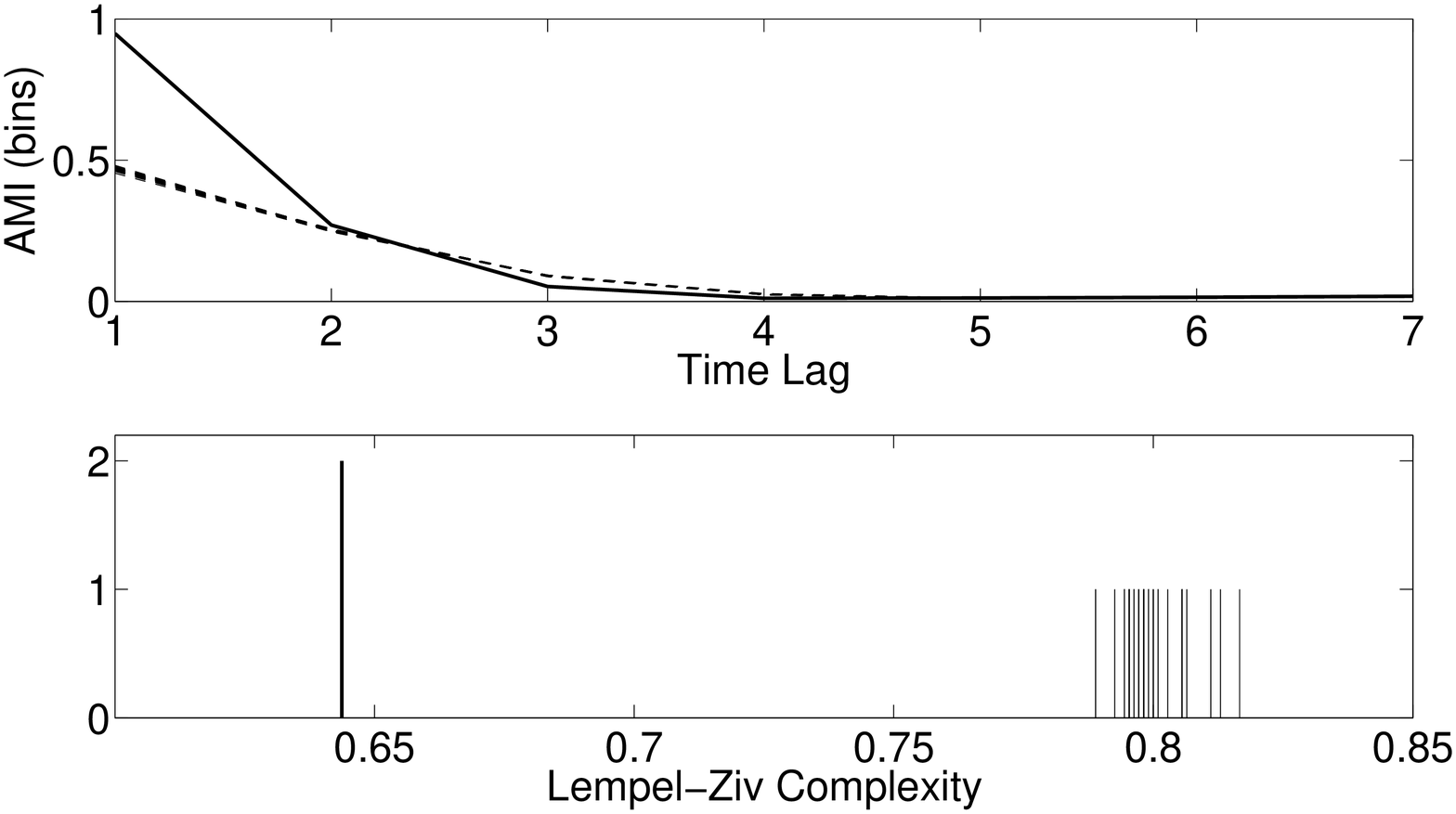}\\
  \caption{AMI and LZ Complexity for one of the signal (continuous line in the top and the longer stem in the bottom)  and the surrogates generated with the SSS algorithm }\label{fig4}
\end{figure}

Fig. \ref{fig3} shows how both algorithms work, the SSS method
randomly destroys the local structures of the data but preserves the
long-term behavior, thus one obtains a realization of a non
stationary stochastic process; while the TFTS method randomly alters
the high frequency components of the signals, whilst preserving the
low frequency components, the surrogates preserve the long-term
behavior of the data, thus obtaining a realization of a non
stationary linear noisy process.\\ Following the procedure stated in
\ref{sec3-d}, we found that the hypothesis addressed by the SSS
algorithm could be rejected, i.e., we found a statistical difference
between the data and the surrogates generated with the SSS
algorithm. This implies that the MER signals possess dynamics. Fig.
\ref{fig4} shows the statical difference between the data and the
surrogates.

\begin{figure}[b]
  \centering
  \includegraphics[width=7cm]{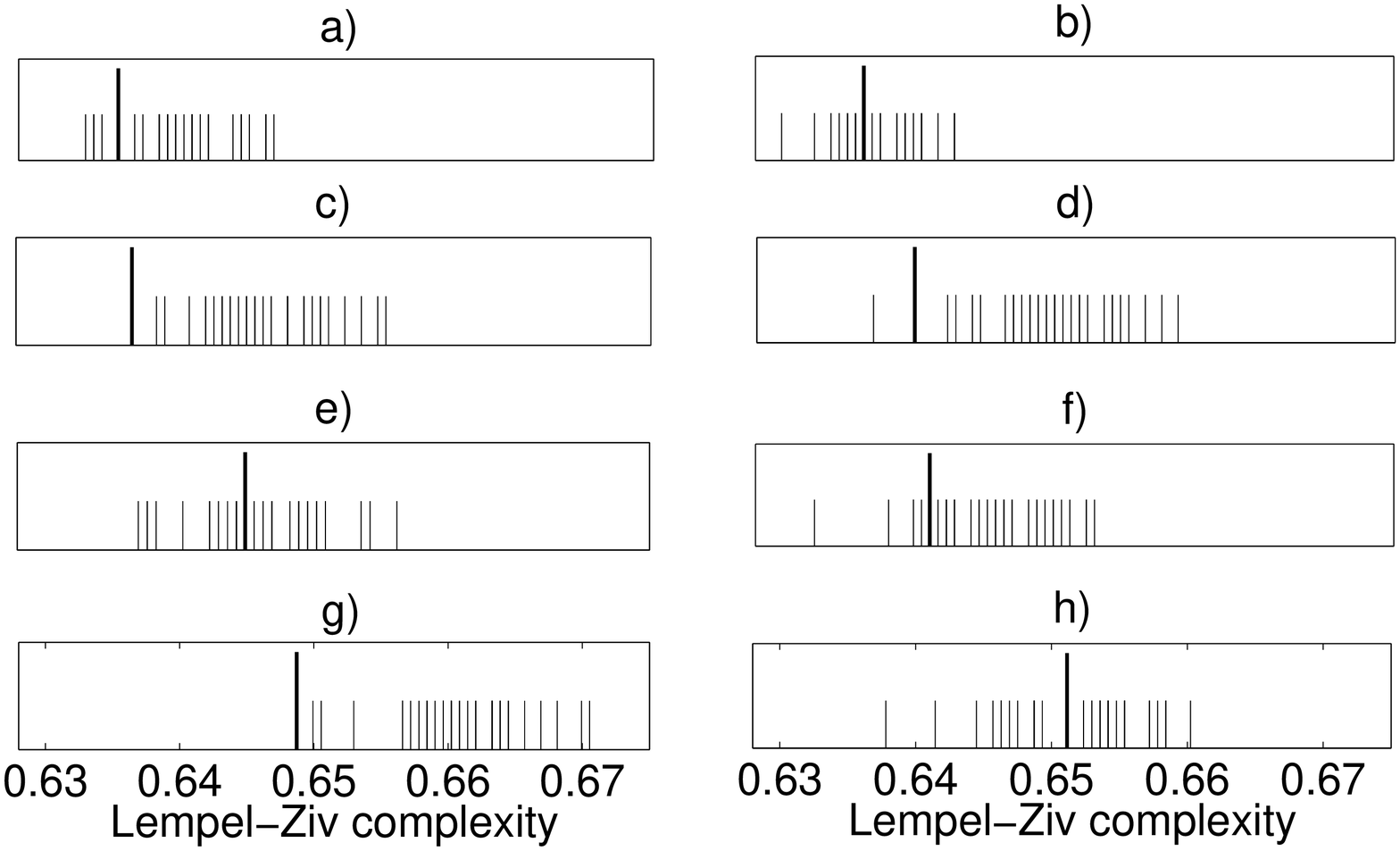}\\
  \caption{Acceptance (a,b,d,e,f,h) or rejection (c,g) of the null hypothesis addressed by the TFTS method using the LZ complexity. }\label{fig5}
\end{figure}

Applying the TFTS method we encountered that almost $27\%$ of the
database rejects the null hypothesis, while the rest of the database
was not able no reject it. First, it is necessary to clarify that
the fact that we were not able to reject the null hypothesis does
not make it true, it just means that the statistical methods found
no difference between the
original data and the surrogates.\\
What is unusual here is that not all the database behaves in the
same way (regarding the null hypothesis) so, we need to seek an
explanation for this phenomenon. If this odd behaviour where caused
by a miss application of the TFTS method, the null hypothesis would
be rejected or accepted by all the signals of the database (in Fig.
\ref{fig4} we present the result of a miss application of the
method). An other possible explanation is that, the 1s window turns
out to be stationary. This is possible but very improbable. To prove
this, we applied a stationarity test proposed in \cite{Borgnat2009}
to the signals
that reject the null, we found that all the signals were non stationary.  \\
Finally, we apply the following procedure: We take the 10s signal
(for this we only used the UPV database) and divided it into 8
sub-segment of 1s (we did not use the first and last second of the
signal), then applied the procedure described in \ref{sec3-d} to
each sub-segment. We found that each signal poses sub-segments that
reject the null and other sub-segments that accept the null. That
is, for some time intervals the signal behaves like the realization
of a non-stationary linear noisy process and for some other it does
not. Fig. \ref{fig5} shows this result for one of the signals using
the LZ complexity as test statistic.
\section{Conclusions}
Through our methodology we proved that the MER signals are
deterministic, this in contrast to what has been guessed by some
authors \cite{Xue,Pinzon}. This implies that there is a dynamic rule
that governs the temporal evolution of the signal. Unfortunately due
to non-stationarity of the signal, there is not a
good method for estimating the dimension of the dynamical system.\\
We found that there are moments in which the MER signal can be
modelled as a realization of a non stationary linear noisy system
and others in which it may not, so we might conclude that
methodologies such as the wavelet transform or the Hilbert-Huang
transform are suited for the analysis of MER signals. We also
encourage researchers not to characterize MER signals through
methodologies developed for time series that behave like i.i.d.
random variables or that are limited to be applied to stationary
processes (e.g. Fourier analysis, non linear dynamics analysis).


\section{Acknowledgments}

We express our sincere appreciation to the Research Center of the
Instituto Tecnológico Metropolitano of Medellín - Colombia within
the framework of the P09225 grant, Universidad Tecnológica de
Pereira and the Créditos Condonables program financed by
COLCIENCIAS. We wish to thank T. Nakamura, for his contributions
during this study. We also appreciate the comments by two anonymous
reviewers.


\addtolength{\textheight}{-3cm}   
%
%
%
%
%
\bibliographystyle{IEEEtran}
\bibliography{bib}
\end{document}